\documentclass[10pt]{article}

\usepackage{amsmath}
\usepackage{amssymb}

\usepackage{cite}

\usepackage{hyperref}

\usepackage{lineno}

\usepackage{microtype}
\DisableLigatures[f]{encoding = *, family = * }

\usepackage{setspace} 

\usepackage{graphicx}
\usepackage{xcolor}
\usepackage[normalem]{ulem}

\usepackage[labelfont=bf,labelsep=period,justification=raggedright]{caption}

\makeatletter
\renewcommand{\@biblabel}[1]{\quad#1.}
\makeatother

\date{}

\begin{document}

\begin{flushleft}
{\Large
\textbf{Behavior of early warnings near the critical temperature in the two-dimensional Ising model.}
}
\\
Irving O. Morales$^{1, 2, 6,\ast}$, 
Emmanuel Landa$^{3, 6}$,
Carlos Calderon Angeles$^{4, 6}$,
Juan C. Toledo$^{1,2, 6}$,
Ana Leonor Rivera$^{1,5, 6}$,
Joel Mendoza-Temis$^{1,2, 6}$,
Alejandro Frank$^{1,2, 6}$.
\\
\bf{1} Instituto de Ciencias Nucleares, Universidad Nacional Aut\'onoma de M\'exico, M\'exico, D.F., M\'exico. 
\\
\bf{2} Centro de Ciencias de la Complejidad, Universidad Nacional Aut\'onoma de M\'exico, M\'exico, D.F., M\'exico.
\\
\bf{3} Instituto de Bioci\^encias, Universidad de S\~ao Paulo, SP, Brazil.
\\
\bf{4} Facultad de Ciencias, Universidad Nacional Autonoma de M\'exico, D.F., M\'exico.
\\
\bf{5} Centro de F\'isica Aplicada y Tecnolog\'ia Avanzada, Universidad Nacional Aut\'onoma de M\'exico, Quer\'etaro, M\'exico.
\\
\bf{6} Laboratorio Nacional de Ciencias de la Complejidad, M\'exico.
\\
$\ast$ E-mail: Corresponding irvingm@nucleares.unam.mx
\end{flushleft}

\section*{Abstract}
Among the properties that are common to complex systems, the presence of critical thresholds in the dynamics of the system is one of the most important. Recently, there has been interest in the universalities that occur in the behavior of systems near critical points. These universal properties make it possible to estimate how far a system is from a critical threshold. Several early-warning signals have been reported in time series representing systems near catastrophic shifts. The proper understanding of these early-warnings may allow the prediction and perhaps control of these dramatic shifts in a wide variety of systems. In this paper we analyze this universal behavior for a system that is a paradigm of phase transitions, the Ising model. We study the behavior of the early-warning signals and the way the temporal correlations of the system increase when the system is near the critical point.


\section*{Introduction}

Complex systems are those in which the agents or elements that compose the system interact non-linearly and in such a convoluted way that it is impossible to describe the behavior of the system in terms of the simpler behavior of its components. Intuitively, it should be easy to define precisely what a complex system is and under which conditions complexity emerges. However, so far there is no general agreement of what complex systems are. In fact, the very definition of complexity is a main research topic\cite{Prokopenko:2009cj,Gershenson:2012ft,Crutchfield:1989hp,Razak-FA:2014uq}. Complex systems share certain general properties which in some fashion describe them: emergence, self-organization, homeostasis, entangled properties on multiple scales, ability to efficiently transmit and process information, etc.\cite{Prokopenko:2009cj,Gershenson:2012ft,Razak-FA:2014uq}. In this work we focus on a property that is apparently common to many complex systems: the existence of critical thresholds\cite{Scheffer:2009sy}. A large number of complex systems display behaviors related to criticality and phase transitions\cite{Sole:1996ng,Bak:1988bx,Landa:2011cz}. When a physical system is in a critical point it acquires unique properties, one of which is scale invariance. Theoretically, at that point the correlation length diverges, which in practice means that the correlation length becomes very large when compared to the scales of interaction of the system. These properties may be the key to the robustness and adaptability of complex systems. One of the main properties that most of complex systems share is the fact that they seem to lie on the boundary between order and chaos\cite{Kauffman:1993zw}. It is in this regime where complex systems acquire robustness and where they can adapt to environmental changes. If the system has excessive order, it tends to be too rigid to react promptly to the pressures of the environment, and is thus incapable of responding to its requirements. It is not capable of evolving. On the other hand, if the system behaves too randomly, then it is too fragile and any environmental perturbation will impair fundamental properties of the system. Living on the edge between order and chaos, complex systems are capable of being adaptive to changes in a robust way. In the physical sciences, this regime between order and chaos is frequently associated with phase transitions and criticality. Although the relation between robustness, adaptability and criticality in complex systems is quite evident, a second way in which the complex systems are related to the phase transition phenomena is through the appearance of catastrophic shifts. It is known that many complex systems have critical thresholds at which the system changes dramatically from one stable state to another. Usually these changes are abrupt and dramatic relative to the long-range time scales of the system. These shifts are driven by small perturbations (again in terms relative to the typical scales of the system), follow paths with hysteresis, and are related to the critical slowing down phenomenon\cite{Scheffer:2009sy}. This kind of transition is present in a large and diverse set of complex systems, which include ecosystems\cite{Carpenter:1999il,deYoung:2008to,Rietkerk:2004qc}, species population\cite{Dai:2012tx}, biodiversity, the global economy\cite{May:2008wu}, physiological systems\cite{Litt:2001wc,McSharry:2003bq,Venegas:2005qp}, climate systems\cite{Kleinen:2003tk}, social systems, etc. Although the particular features of these systems are quite diverse, all these transitions display characteristic signals which in principle are independent of the particularities of the system. In other words, the dynamics of systems near critical points exhibit universal properties. There has recently been an increased interest in understanding how a complex system behaves in the vicinity of catastrophic shifts, in part to predict and possibly control the timing and evolution of such transitions. Usually early warnings and critical transitions are studied either in model systems, for which it is possible to describe the behavior of the system in terms of a mathematical model, or in actual complex systems in which the description is based in a rich collection of data, but where a full mathematical description is not available\cite{Scheffer:2009sy}. In this work we focus on a very well known physical system, the Ising model \cite{Ising:1925qd}. We choose this model because the phase transition associated to it is very well known and the presence of critical slowing down is a well studied phenomenon\cite{Wang:1993hl}. For the study of biological systems, the Ising model has been applied to the study of human brain\cite{bib:Marinazzo_2014, bib:Das_2014, bib:Kitzbichler_2009}, cancer \cite{bib:Torquato_2011}, protein folding\cite{bib:Henry_2013}, ion channels \cite{bib:Liu_1993}, statistical genetics \cite{bib:Majewski_2001} and cardiac activation\cite{bib:Rice_2003}. The Ising model has also been used successfully to study collective phenomena in social systems\cite{Castellano:2009}, specially in order to simulate crowd dynamics and opinion formation. It is a system on which we can delicately control the parameters and the location of the critical point, but that is not trivial in terms of its dynamics, i.e., it is not possible to fully understand the system in terms of dynamical equations in a trivial way. The main goal of the present work is to analyze the dynamics of the Ising model in terms of the classic early warning properties described in literature.\\

\section*{Methods}

\subsection*{The Ising model}

\begin{figure}
\begin{center}
\includegraphics[height=\textheight]{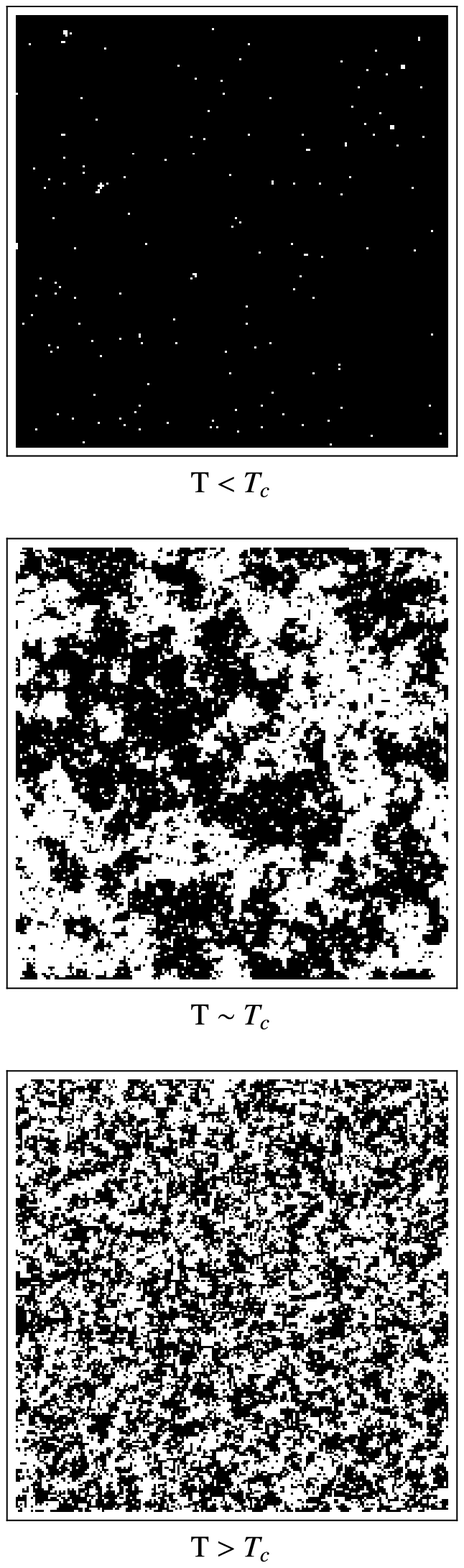}
\caption{
{\bf Spatial configurations in the Ising model.}  Typical spatial configurations for a 2-dimensional Ising model. Three regimes are shown: a) $T < T_c$, b) $T\approx T_c$ and c) $T > T_c$. Black squares represent spins with $\sigma=+1$ and white one correspond to $\sigma=-1$.  
}
\label{fig1}
\end{center}
\end{figure}

The Ising model is a statistical physics model for ferromagnetism. It is paradigmatic both for systems in which cooperative phenomena play an important role and for the study of physical phase transitions. The definition of the system is very simple. Consider a lattice of $N$ sites with a spin state $\sigma$ defined on each site. We let each of the spins take one of two possible orientation values, denoted by $\sigma = \pm 1$. There are thus $2^N$ possible configurations of the system. Each of these spin sites interact with its nearest neighbors with an interaction energy given by 
\begin{equation}
H(\sigma)=\displaystyle-\sum_{i,j}J_{ij}\sigma_i\sigma_j-\mu \sum_{i=1}^N B_i\sigma_i
\end{equation}
where the first summation runs only through neighboring spins and $J_{ij}$ represents the coupling strength between spins $i$ and $j$. If this coupling is positive then the neighboring spins will tend to align parallel to each other, since this minimizes the energy. $B_i$ represents the external magnetic field acting on site $i$ and $\mu$ is the magnetic moment. In this work we focus only in the case where $J_{ij}=$ constant and $B_i = 0$ (no external magnetic field). The probability that the system is in a given configuration depends on the energy of the configuration and the value of the parameter $T$, which is identified as the temperature of the system. This probability is given by the Boltzmann distribution
\begin{equation}
P_\beta(\sigma)=\frac{e^{-\beta H(\sigma)}}{Z_\beta}
\end{equation}
where $\beta=(kT)^{-1}$, $k$ is the Boltzmann constant and the normalization $Z_\beta$ is the partition function. It is possible to measure the order present in the system through the total magnetization, defined as
\begin{equation}
M=\displaystyle \frac{1}{N}\sum_{i=1}^N\sigma_i
\end{equation}
A well-known fact is that if it is defined on a 1-dimensional lattice, in which each spin has only two nearest neighbors, the system will have no phase transition. However, for lattices in 2 or more dimensions the system goes through a phase transition when $T$ is equal to a critical value $T_c$. Below the critical value, the system undergoes spontaneous magnetization and all the spins tend to align towards either the $+1$ state or the $-1$ state. For temperatures higher than $T_c$, the system becomes paramagnetic, where the total magnetization of the system is zero on average. The presence and size of clusters of equally aligned spins is also markedly different in these two regimes: when $T$ is lower than $T_c$, large resilient clusters form, while above $T_c$ only small clusters can survive momentarily. If the temperature is high enough, all the clusters are completely destroyed. In the critical point ($T=T_c$), however, clusters are continually formed and destroyed  in a wide range of scales, with the distribution of cluster sizes following a power law. Figure \ref{fig1} shows a typical spatial configuration for each of the three regimes of temperature for a 2-dimensional system. Black squares represent spins with $\sigma=+1$ and white ones represent those with $\sigma=-1$.\\

While we are interested in the dynamics of the Ising model, the model described thus far does not incorporate dynamics since there is no kinetic term in the Hamiltonian. A kind of artificial dynamics can be imposed on the system through Monte Carlo simulations using the Metropolis algorithm\cite{Metropolis:1953qr}. This algorithm iteratively generates successive spin configurations. While these configurations do not represent the time evolution of a system of spins, it is possible to associate them to such a system in contact with a heat reservoir through Glauber dynamics\cite{Glauber:1963zf}. We will analyze the successive configurations obtained with the simulation assuming that they represent the evolution of a correlated system. We are not interested in the Ising system as a model for ferromagnetism, but rather as a nearest neighbor interaction system in which two competing effects are acting: on the one hand, the short-range interactions between neighbors, while on the other the stochastic fluctuations caused by temperature. We are thus not interested in recovering the usual thermodynamic properties through the simulation. While the critical slowing down phenomenon near the critical transition has been previously seen with the Metropolis algorithm\cite{Wang:1993hl}, the focus has always been on trying to avoid it since the main goal of Monte Carlo simulations is to obtain independent configurations of the system. The presence of critical slowing down is an indication that the successive configurations obtained through the simulation are correlated, i.e., not independent. Usually, the critical slowing down is estimated through relaxation times and it is considered a deficiency rather than a feature of the system. It is thus typically used to determine how many simulation steps need to be skipped in order to obtain independent configurations. In this work we take an opposite view and consider critical slowing down as a feature of the system which can be used to detect criticality. Based on this we simulate the Ising model through the Metropolis algorithm and consider correlation between successive configurations as an indication of early warnings of an oncoming critical threshold.\\

\subsection*{The Metropolis algorithm}

The goal of the Metropolis\cite{Metropolis:1953qr}  Monte Carlo simulation is to generate a large number of different and independent configurations of the system in order to have enough statistical sampling to estimate average values of thermodynamic properties. In this algorithm, new configurations are generated from a previous state using a transition probability which typically depends on the energy difference between the initial and final states. The exact form of the probability comes from considering detailed balance in the master equation for the transition probabilities. Following the Boltzmann distribution, the probability of the system being in a state $n$ is:
\begin{equation}
P_n=\frac{e^{-E_n/kT}}{Z}
\end{equation}
where $E_n$ is the energy of the state, $k$ is the Boltzmann constant, $T$ is the temperature and $Z$ is the partition function. The transition probability from state $n$ to state $m$ is then given by
\begin{equation}
P_{n\rightarrow m}=\exp[{-\Delta E/kT}]
\end{equation}
where $\Delta E=E_m - E_n$.

Given a previous state of the system, defined by a specific set of values for all spin sites $\sigma_i$, the algorithm proceeds as follows:

\begin{enumerate}
\item Randomly choose a site $i$;
\item Calculate the change in energy $\Delta E$ if spin site $i$ were to be flipped;
\item If $\Delta E$ is negative, then flip the spin of site $i$. If on the other hand $\Delta E$ is positive, generate a uniformly distributed random number between 0 and 1 and flip the spin only if this random number is less than $\exp(-\Delta E/kT)$;
\item Choose another site and go back to step 1.
\end{enumerate}

\begin{figure}
\begin{center}
\includegraphics[height=\textheight]{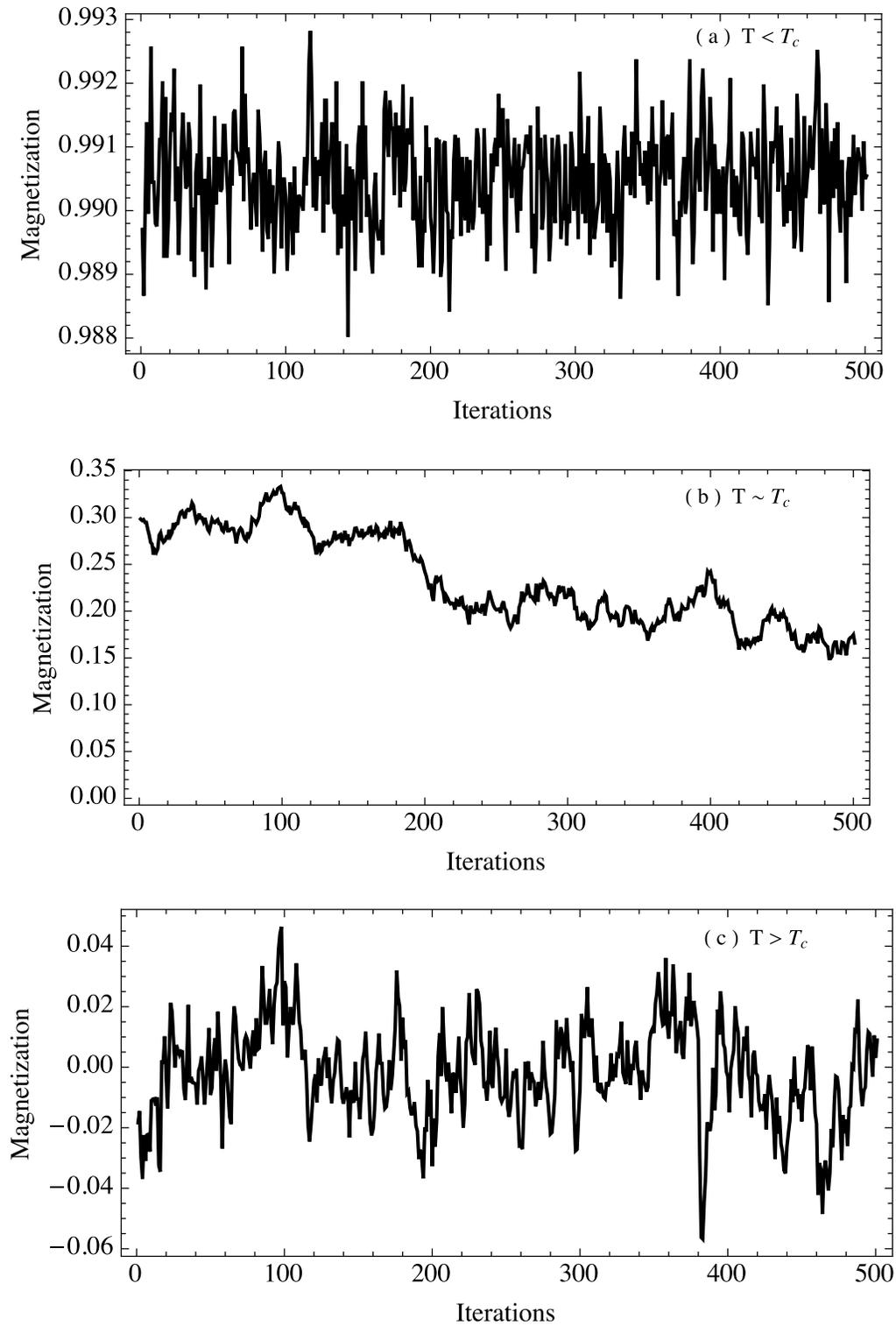}
\caption{
{\bf Total magnetization as a function of time in the Ising model.}  Typical behavior of the total magnetization time series in a 2-dimensional Ising model. Three regimes are shown: a) $T < T_c$, b) $T\approx T_c$ and c) $T > T_c$. It is important to notice the change of scale between plots.
}
\label{fig2}
\end{center}
\end{figure}

An iteration or a simulation time unit has passed when every spin in the system has had a chance to flip. We randomly choose the order in which the spin sites are selected for the process described above. It is common that when the system temperature is near the critical point $T_c$, relaxation times are computed in order to skip several iterations and avoid correlations between successive configurations. In this analysis we keep all the iterations because we are precisely interested in the correlations in the system and how these correlations behave as $T$ approaches $T_c$. Particularly, we analyzed an Ising system defined over a 2-dimensional square lattice so that each spin has 4 nearest neighbors. We choose the units of the system such that $T_c \approx 2.27$ which means that $k=1$ and $J_{ij}=1$. The lattice has a finite size of $100 \times 100$ sites with periodic boundary conditions. Although the finiteness of the lattice can affect the value of $T_c$, we are not interested in simulating the system precisely at the critical point but rather in wider regions near criticality. We simulate the system for three temperature regimes: $T < T_c$, $T \approx T_c$ and $T > T_c$. For temperatures lower than $T_c$ we choose as initial condition a configuration where all the spins are aligned with the $+1$ state. We do this in order to avoid as much as possible the metastable configurations typical of this regime (metastable states are described later). For both regions where $T \approx T_c$ and $T > T_c$, we choose an initial condition where all the initial spin states are randomized, so that the average total magnetization is $M = 0$. We sampled the system from $T = 1.42$ to $T = 3.12$, changing the temperature in intervals of $\Delta T = 0.05$. For each of these temperatures we run an ensemble of 1000 simulations of 5000 iterations each (where the first 1000 iterations are discarded in order to remove the initial transient). We then computed the total magnetization of the system as a function of time and analyzed the properties of the magnetization temporal fluctuations considering them as a time series. It is important to note that we are not dynamically changing the temperature; its value is fixed for every run of the simulation. Thus our system is not being driven towards the critical threshold. This can also be thought as the rate of change of temperature being very small compared to the dynamic scale of the system. Under this assumption, we can consider that we are sampling separate instants of this very slow transition through the critical threshold. Figure \ref{fig2} shows a typical magnetization time series for each of the three temperature regimes. For each temperature value we computed the ensemble average of several parameters that can function as early warnings in order to better understand how the system behaves when it is near the critical region.

\subsection*{Early Warnings}

The variety of early warning signals that have been proposed in the literature is overwhelming. There recently has been an effort to summarize, compare and contextualize the different techniques that have thus far been explored\cite{Scheffer:2009sy,Dakos:2012ht}. It is possible to broadly classify early warning signals in two main categories: the {\em metric-based} indicators which essentially detect subtle changes in the statistical properties of the time series, and the {\em model-based} indicators which detect changes in the time series dynamic fitted by a reasonable model. In this work we will focus only on the former approach. Among the {\em metric-based} estimators we can distinguish two main lines of thought.

The first one is based on the non-stationarity of the time series. If we think of the time series as a stochastic process, it is possible to estimate the probability distribution of events. Theoretically, if the system approaches a critical point then due to critical slowing down several moments of the probability distribution will change. The second moment of the distribution, the variance, will diverge because near a critical threshold a system recovers very slowly from perturbations, which in principle allows the system to drift across the boundaries of different states \cite{Carpenter:2006hb}. Depending on the particularities of the system, it is possible that the fluctuations become asymmetric if the system approaches configurations with an unstable equilibrium. This will produce changes in the third moment of the distribution, the skewness \cite{Guttal:2008hw}. Changes in the fourth moment of the distribution are also possible, because near criticality the system will visit extreme states more often \cite{Dakos:2012ht}.

The second approach in the family of {\em metric-based} estimators is to analyze changes in the memory of the time series via its temporal correlations. Changes in the temporal correlation of events are strongly linked with the situation in which the system approaches a tipping point. If the system recovers slowly from perturbations, as happens near the critical threshold, then it is expected that the temporal correlations of the system for short time scales will increase. This effect can be calculated by means of the autocorrelation function for small time lags $\tau$, specially through the autocorrelation at lag 1, $C(\tau=1)$\cite{Dakos:2008lp}. The correlations for long time scales are also modified when the system approaches criticality. These correlations are associated with slow oscillations in the fluctuations and with long range memory effects in the system. This kind of effects are visible when an analysis for all the temporal scales is performed. Some of the effects most commonly used early warnings are the Power Spectral Density (PSD) analysis\cite{Kleinen:2003tk} and the Detrended Fluctuation Analysis (DFA)\cite{Peng:1994ye}. The PSD is connected to the amount of correlations present in the system. Particularly, it is well known that the PSD of scale invariant series obey a power law\cite{Bak:1988bx,Landa:2011cz}. It is important to mention that in this work we have focused on temporal early warnings. However, given the rich spatial behavior of the Ising model, it would also be interesting to study early warnings in the spatial domain \cite{Dakos:2011,Kefi:2014}. It is our intention to analyze this kind of early warnings and their relation with the magnetization cluster patterns formed when the Ising system approaches the critical region in a future work.

\section*{Results and Discussion}
We carried out simulations of the Ising model dynamics for a range of temperatures. We followed the total magnetization of the system through time as the simulation evolved, and in this way we constructed a magnetization time series for each simulation. We repeated the experiment 1000 times for each temperature in order to estimate the ensemble behavior of the system. For each temperature we computed the ensemble average of the early-warning signals described previously, the first four moments of the probability distribution, the autocorrelation at lag 1 and the power spectral density.\\

\begin{figure}
\begin{center}
\includegraphics[width=\textwidth]{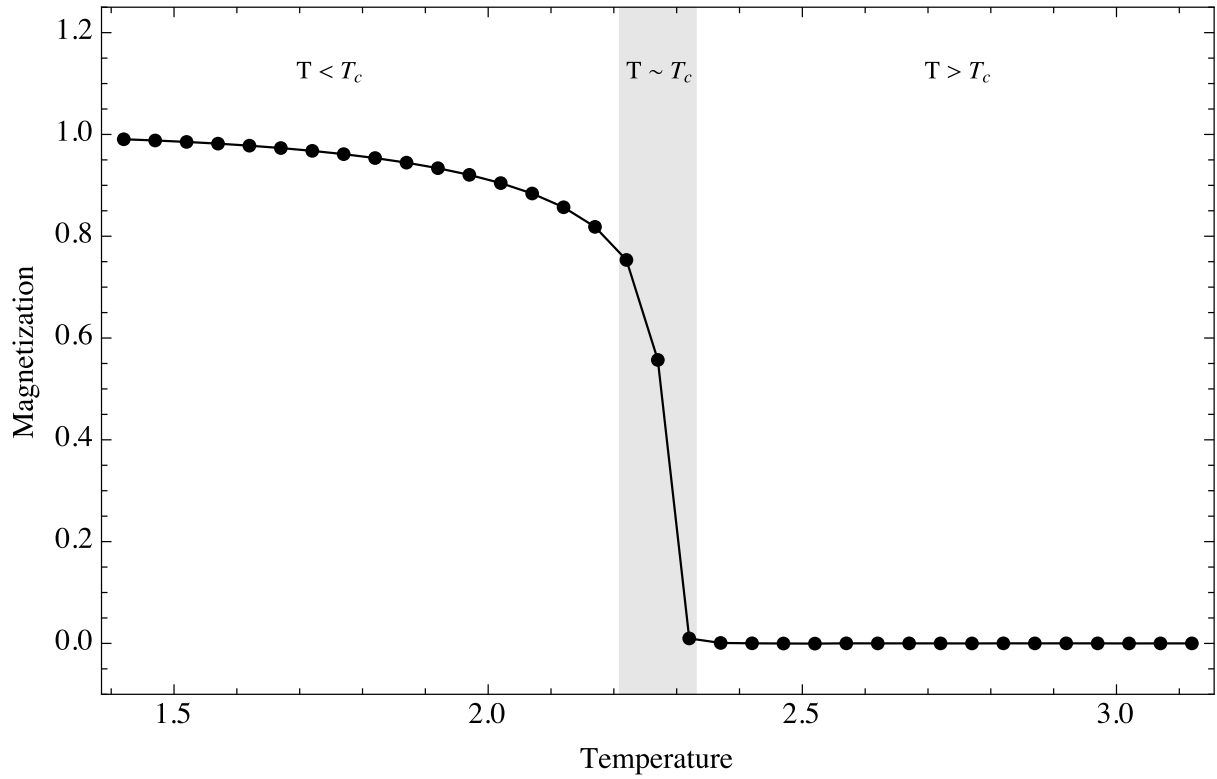}
\caption{
{\bf Temporal mean as a function of temperature.} Ensemble behavior of the mean as a function of temperature. The mean corresponds to the total magnetization of the system. Three regimes are shown, $T < T_c$, $T\approx T_c$ and $T > T_c$. Note that the mean can also approach $-1$ at low temperatures; we only show here the positive values.
}
\label{fig3}
\end{center}
\end{figure}

The first moment of the distribution is the mean and corresponds to the average magnetization of the system. Figure \ref{fig3} shows the ensemble average for the magnetization as a function of temperature (it is important to mention that for low temperatures the global magnetization can also converge towards $-1$; we have thus in general used only the positive values). It is clear that our simulation follows the usual behavior of an Ising model. The three regimes can be clearly identified in the simulations. a) For $T>T_c$, the stochastic contribution to the fluctuations dominates the neighbor spin interactions. In this regime clusters of magnetization are very small, and practically all spin sites change orientation independently of the state of the neighboring spins. The system is governed by stochastic fluctuations and any attempt to transmit information through the system will fail, since correlations are quickly destroyed by the high stochasticity of the system. b) On the other hand, for $T<T_c$, the main contribution to the dynamics comes from the short range interaction between neighbors. In this regime large clusters of magnetization are formed, which is shown by the fact that magnetization does not quite reach a value of 1. Again, any attempt of transmit information fails, essentially because any  spin flip is promptly outweighed by the dominating neighbor interaction. c) Finally, at the critical point, $T=T_c$, the two opposing effects are in balance. As we mentioned before, in this regime magnetization clusters of various sizes are formed and the corresponding size distribution of the clusters follows a power law, which is a very strong indication that the system is spatially scale invariant. The effects of perturbations in this regime are stable and robust. \cite{jeldtoft:1998} succinctly summarized this phenomenon as follows: ``For all other temperatures, one can disturb the system locally and the effect of the perturbation will influence only the local neighborhood. However at the transition temperature, the local distortion will propagate throughout the entire system. The effect decays only algebraically rather than exponentially. Although only `nearest neighbor' members of the system interact directly, the interaction effectively reaches across the entire system. The system becomes critical in the sense that all members of the system influence each other." The fundamental emergent property of the system near criticality is its capacity to transmit information over scales comparable to the entire scale of the whole system. While the average magnetization of the system is zero, it is located precisely at the frontier between the zero magnetization regime and the magnetized one.\\

\begin{figure}
\begin{center}
\includegraphics[width=\textwidth]{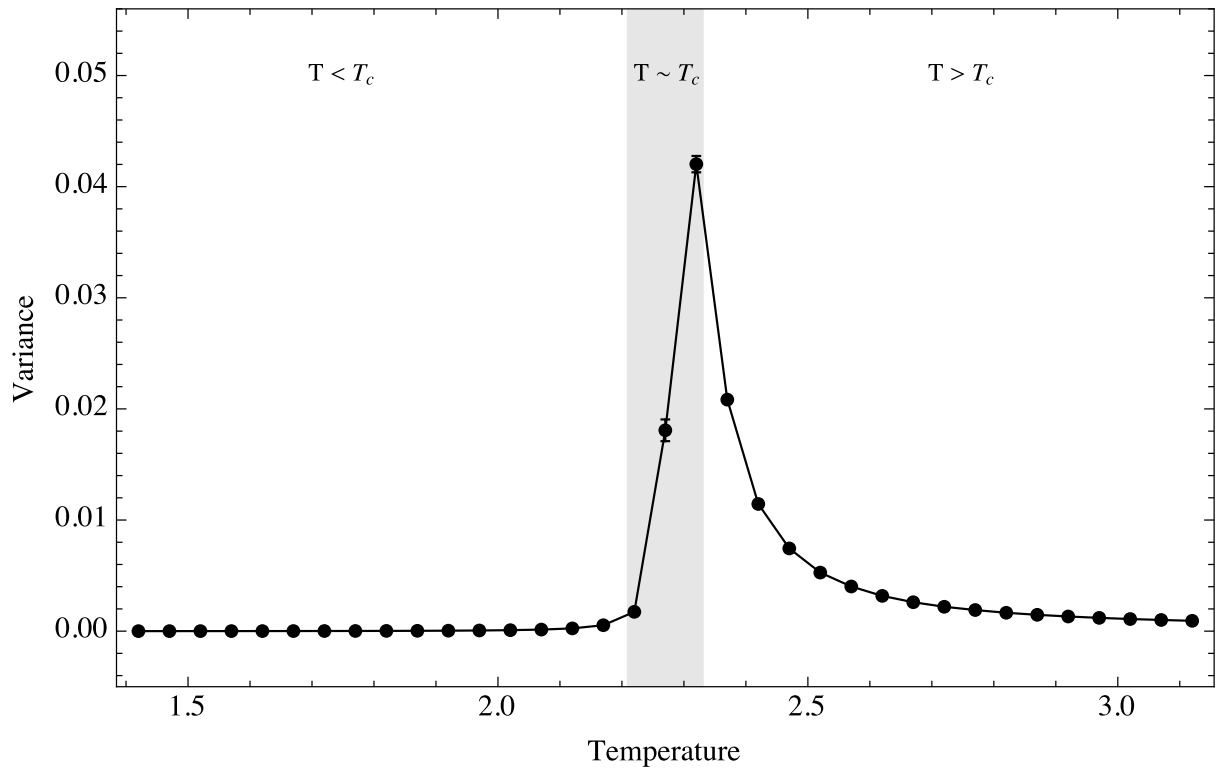}
\caption{
{\bf Temporal variance as a function of temperature.} Ensemble behavior of the variance as a function of temperature. Three regimes are shown, $T < T_c$, $T\approx T_c$ and $T > T_c$.  
}
\label{fig4}
\end{center}
\end{figure}

Changes in the higher moments of the distribution constitute important early warnings. Figure \ref{fig4} shows the ensamble behavior of the temporal variance of the magnetization  and the way it changes as a function of temperature. There are two ways to approach the critical value: either reaching it from temperatures below $T_c$, or coming from higher temperatures. It is evident that the response of the system is different in these two cases. At the critical point the variance is maximal, but its increase is quite different depending on which side of the critical point the system is coming from. In the high temperature regime the variance grows smoothly as the temperature diminishes. In this regime the variance will constitute an excellent early-warning signal, for its changing behavior will indicate that the system is approaching a critical region even if we are not aware beforehand of the existence of a criticial point. On the other hand, for the low temperature regime the variance increases abruptly, so it is difficult to use the variance as an early-warning signal when reaching the critical temperature from below. It is also interesting that the ensamble standard error of the mean (defined as the standard deviation divided by the square root of the sample size and shown with error bars in the figures) becomes larger when the system is near the critical point.\\

\begin{figure}
\begin{center}
\includegraphics[width=\textwidth]{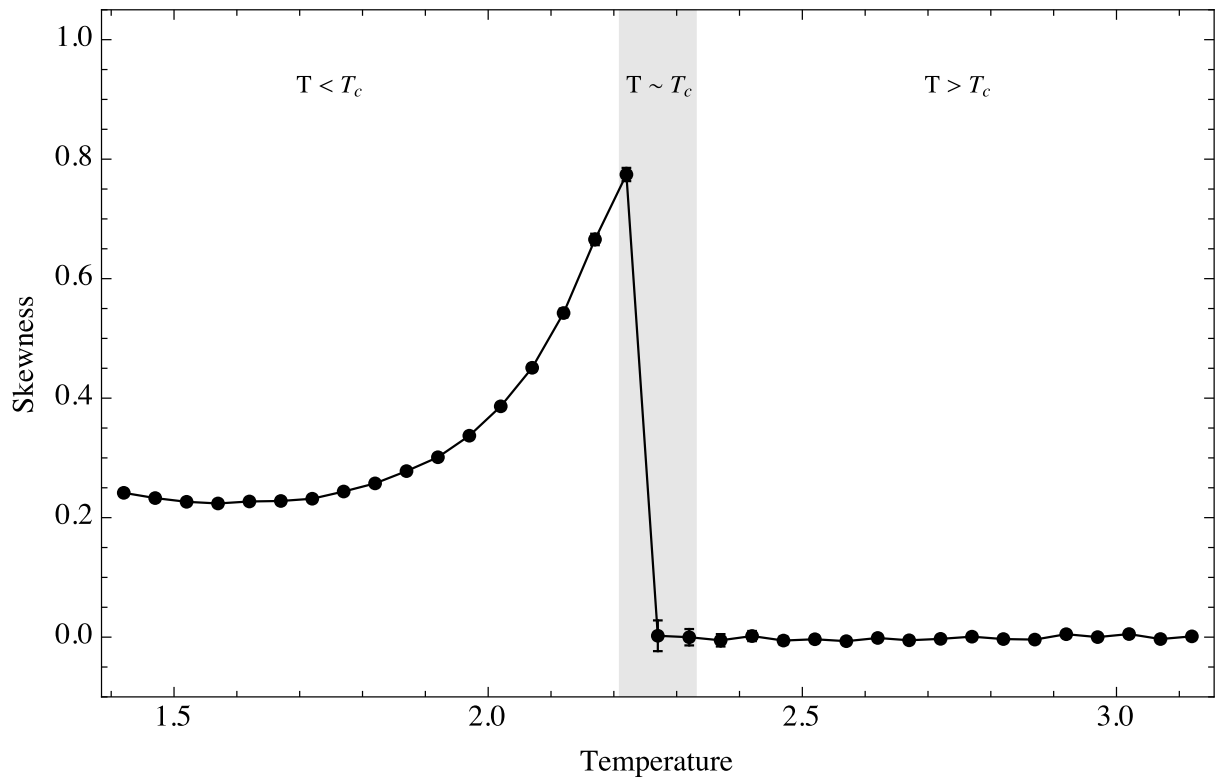}
\caption{
{\bf Absolute values of temporal skewness as a function of temperature.} Ensemble behavior of the skewness as a function of temperature. Three regimes are shown, $T < T_c$, $T\approx T_c$ and $T > T_c$.  
}
\label{fig5}
\end{center}
\end{figure}

The third moment of the distribution, the skewness, is shown in Figure \ref{fig5}. This moment is related to the asymmetry of events in the time series, and we expect that this asymmetry to appear only in the low temperature regime. The reason for this is that in this regime the system has the possibility of becoming trapped in a meta-stable state where the average magnetization is lower than the expected value. These meta-stable states are produced when instead of one large cluster of magnetization, two stable clusters with opposite alignments are formed. A large perturbation is required for the system to spring out of this state. If the temperature is too low, the stochastic perturbations are too weak compared to the neighbor interactions and the system can spend a long time in meta-stabilty. However, near the critical point the stochastic perturbations are large enough to carry the system away from such meta-stability. This effect increases the skewness of the probability distribution. We thus conclude that skewness is an effective early-warning signal if the system is going towards the critical point from below. When approaching $T_c$ from higher temperatures it is not possible to produce asymmetries in the distribution and thus the skewness is not sensitive to the approaching criticality.\\

\begin{figure}
\begin{center}
\includegraphics[width=\textwidth]{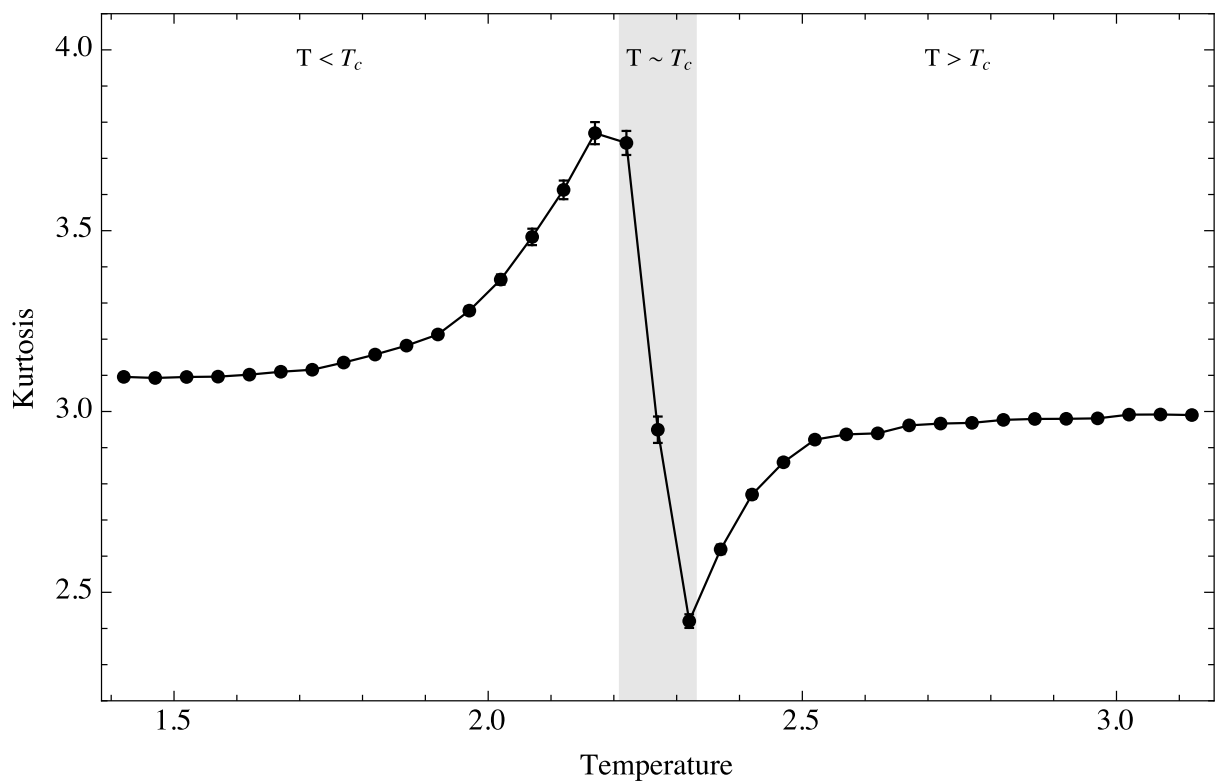}
\caption{
{\bf Temporal kurtosis as a function of temperature.} Ensemble behavior of the kurtosis as a function of temperature. Three regimes are shown, $T < T_c$, $T\approx T_c$ and $T > T_c$.  
}
\label{fig6}
\end{center}
\end{figure}

Figure \ref{fig6} shows the kurtosis of the distribution. It is clear that this moment is a good early warning for this system regardless of the direction in which the system approaches criticality. Kurtosis behaves quite symmetrically and increases smoothly when coming from either direction. The only difference is whether the distribution is more strongly peaked or not than the reference normal distribution (which has a kurtosis of 3). When approaching criticality from lower temperatures we have a strongly peaked or leptokurtic distribution, while when doing so from higher temperatures we have a flattened or platykurtic distribution.\\

\begin{figure}
\begin{center}
\includegraphics[width=\textwidth]{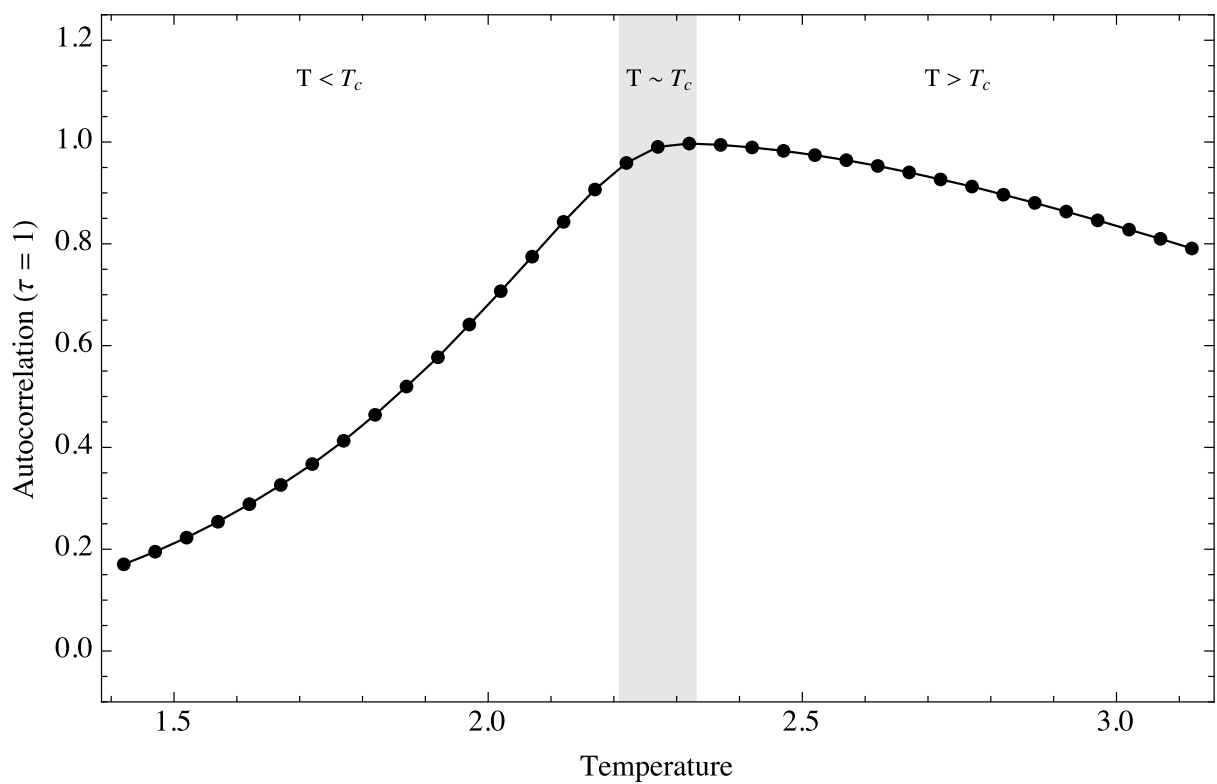}
\caption{
{\bf Temporal auto correlation at lag 1 as a function of temperature.} Ensemble behavior of the autocorrelation function for lag $\tau = 1$ as a function of temperature. Three regimes are shown, $T < T_c$, $T\approx T_c$ and $T > T_c$.  
}
\label{fig7}
\end{center}
\end{figure}

\begin{figure}
\begin{center}
\includegraphics[width=\textwidth]{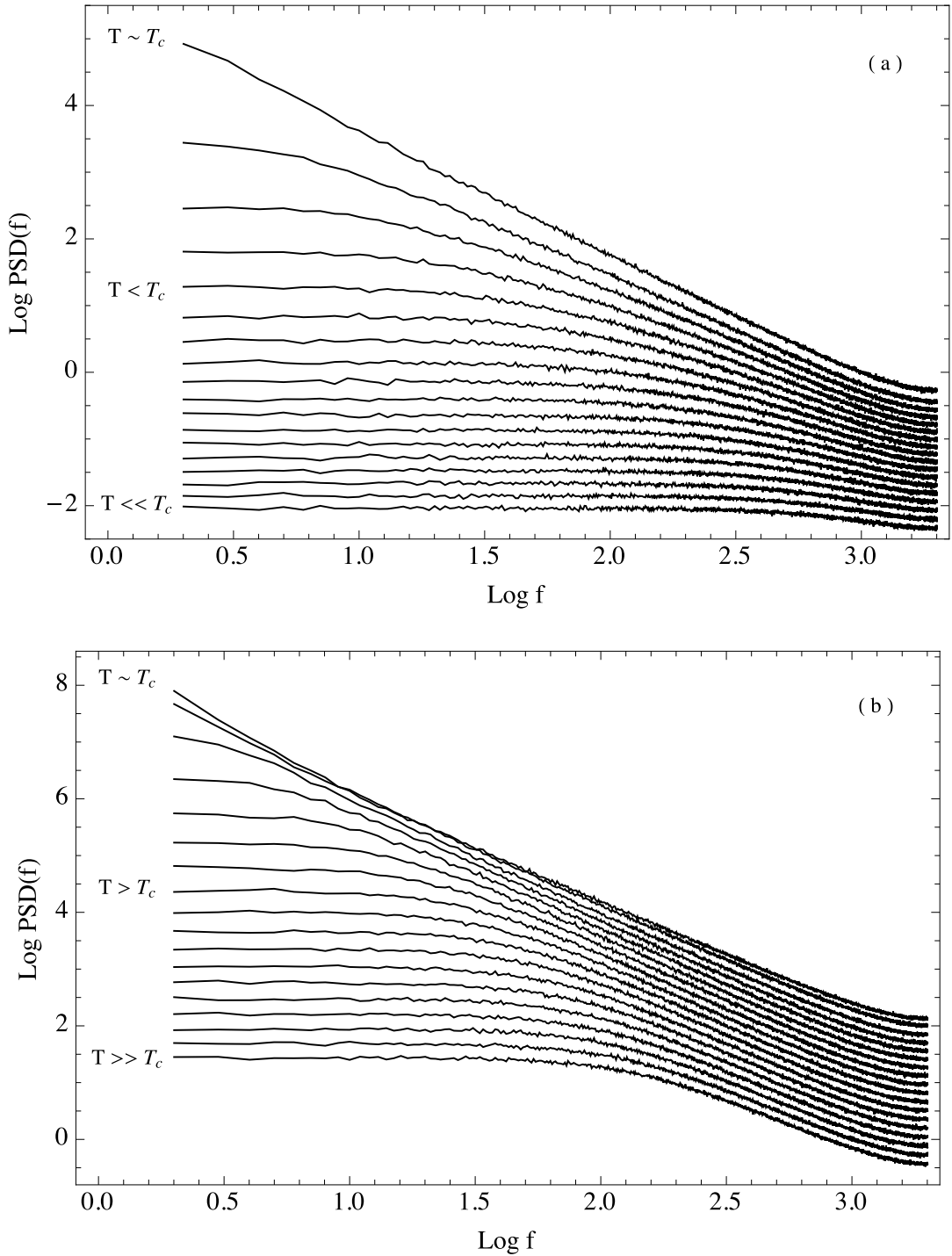}
\caption{
{\bf Power Spectral Density as a function of temperature.} Ensemble behavior of the Power Spectral Density as a function of temperature. Panel (a) shows the behavior of the PSD for temperatures $T \le T_c$. Temperature increases from bottom to top, with $T_c$ corresponding to the topmost curve. Panel (b) shows the behavior of the PSD for temperatures $T \ge T_c$. Temperature increases from top to bottom, with $T_c$ corresponding to the topmost curve.
}
\label{fig8}
\end{center}
\end{figure}

In terms of correlations, it is well known that simulations of the Ising model with the Metropolis algorithm display critical slowing down. It is expected that early warnings based on correlation estimations are useful for this system. Figure \ref{fig7} shows the ensemble behavior of the autocorrelation at lag $\tau = 1$ as a function of temperature. As expected the autocorrelation of the system for very short temporal scales increases when the temperature approaches the critical value $T_c$. The increase is not symmetrical; it is faster when the system approaches the critical value from lower temperatures. However, in both cases it is possible to use this information as an early warning. The autocorrelation is almost 1 for temperatures near the critical value, which means that the system configuration of a given iteration is highly dependent on the previous configuration from which it was obtained. The increase of variance and short term correlations is related to the form of the potential driving the dynamics of the system. They are different if we approach the critical threshold from lower temperatures than from higher temperatures because the form of the potential, basically the way in which it flattens out, is different in these two regions. For this short time scale the system exhibits a clear memory effect. While the short term autocorrelation function is a well known early-warning signal, long term correlations are a better representation of the interesting properties of a critical system. The autocorrelation function is related with the PSD through the Wiener-Khinchin theorem, provided that the time series is a stationary random process\cite{Champeney:1989}. In order to explore the behavior of the long range correlations present we analyzed the PSD of the system. Figure \ref{fig8} shows the memory effect for the whole range of scales. In panel (a) we can observe the evolution of the PSD for temperatures smaller than the critical value. The temperature increases from bottom to top, with the critical value corresponding to the topmost curve. Panel (b) shows the corresponding PSD evolution for the high temperature regime. Again, the topmost curve corresponds to the critical temperature, and this time temperature increases from top to bottom. In both panels the PSD curves for the different temperatures have vertically shifted for clarity. It is clear that a power law appears in the PSD at the critical point. Power laws have been previously connected to criticality, specially with temporal scale invariance\cite{Landa:2011cz,Bak:1988bx}. As we have mentioned, at the critical point the Ising model exhibits spatial scale invariance as well as fractal structure in the sizes of the magnetization clusters formed by the system. It is remarkable that the system also displays temporal scale invariance. Temporal scale invariance means that the time series is statistically the same at all temporal scales. This property is related to long range correlations and long range memory in the system\cite{Landa:2011cz}. As soon as the system's temperature departs from the critical value, either to lower or higher temperatures, the low frequency part of the PSD flattens out. A flat PSD is characteristic of an uncorrelated system, where the fluctuations are white noise. We observe that the flat region of the PSD becomes wider as temperature gets further away from the critical value. For temperatures that are very far from the critical one, we can expect that the PSD will flatten out for all frequencies. For every temperature, there is a crossover frequency at which the power law changes from a power law to the flat profile, an uncorrelated system. This crossover is connected to the temporal scale at which the long range correlations cease to be relevant. States of the system that are separated by a time interval equal or greater than this scale should thus be independent. This behavior is known in the early warning literature as \textit{spectral reddening}\cite{Kleinen:2003tk}. It is a property of the PSD which appears when the system approaches a critical threshold. However, in order to be used as an early warning signal it must first be quantified. The way in which this is usually done is through the spectral exponent or through the spectral ratio between low and high frequencies. However, there are systems for which the power law behavior appears only near the critical threshold, the Ising model being a prime example. We thus consider that the application of the spectral exponent is limited. The crossover frequency where the power spectrum behavior changes is directly related to the range of the temporal correlations, in turn related with the temporal scale at which the system becomes independent from its past states. When a system is near a critical transition, the correlation length increases, and scale invariance appears. In this state the crossover frequency appears at a scale comparable to the scale of the system. Thus, by comparing the scale at which the crossover occurs to the scale of the system, it is possible to estimate how far the system is from a critical transition. It is important to notice that this split in the PSD profile has been reported in other systems in which criticality is a desired property and the appearance of the the two-part spectrum is considered as an indication that the system is losing the properties associated with criticality, i.e., robustness and adaptability. For instance, in physiological time series, and in particular for cardiac interbeat intervals, the PSD exhibits a power law when the heart of the subject is deemed healthy\cite{Kobayashi:1982so,Appel:1989wk}. The underlying hypothesis is that the heart is a complex system that has evolved to be both robust and adaptable, and it has achieved this by approaching spatial and temporal scale invariance. One can think that these properties allow the heart to be resilient against environmental changes while at the same time being able to adapt to the typical efforts to which a heart is subjected. This resilience and criticality are lost over time and due to illness, and it has been reported that when this happens the PSD stops being a power law and a two part spectrum appears\cite{Iyengar-N:1996pd,Lipsitz:1990zg}.\\

The fact that criticality and critical transitions are not the same spawns an interesting question: what is the connection of the critical point in the Ising model with bifurcations? It is certainly possible to analyze the Ising model in terms of a bifurcation. By considering a mean field approximation it is possible to neglect fluctuations in the Ising model. Under this assumption the average magnetization satisfies the following rate equation:
\begin{equation}
\frac{dM(t)}{dt} = -M(t) + \tanh[\beta M(t)] 
\end{equation}
which can be expanded as:
\begin{equation}
\frac{dM(t)}{dt} = -(\beta_c - \beta) M(t) - \frac{1}{3}(\beta M(t))^3 + ...
\end{equation}
This expression is precisely the normal form of a supercritical pitchfork bifurcation, which is typical in physical problems that have symmetry. Under this view it is clear that Figure \ref{fig3} can be thought as the upper half of a pitchfork bifurcation diagram (if we imagine the symmetric negative magnetization branch), and that the stable fixed point at $M = 0$ for $T > Tc$ becomes an unstable fixed point when $T$ goes below $Tc$. The negative derivative of the free energy can be related to an effective force driving the magnetization dynamics. Expanding the free energy as a power series in the magnetization gives:
\begin{equation}
F(M) = C + \frac{1}{2}(\beta_c - \beta)M^2 + \frac{1}{12} \beta^3 M^4 +...
\end{equation}
which is known as the Landau expansion. Below the critical temperature the free energy has two minima and one maximum, while for temperatures greater than the critical value there is a stable minimum at $M = 0$. According to this analysis, the system is bistable for temperatures lower than the critical value. When the system acquires a nonzero magnetization the symmetry of the system has been broken.

\section*{Conclusions}

In this work we have analyzed the behavior of the Ising system in search of the most common early-warning signals. We have selected the Ising model because it is a far from trivial example of a system in which a very well known critical point is present. When the system is near the critical point it exhibits the critical slowing down phenomenon and our numerical experiments demonstrate the deep connection of this behavior with several early-warning signals in the evolution of the magnetization time series. We concentrate on early-warning signals related to the statistical properties of the time series probability distribution. We have shown that the change in these properties is asymmetrical, i.e., it is not the same when the critical point is approached from different directions. This effect restricts the applicability of some of the statistics-based early-warning signals, depending on the system at hand. On the other hand, we also analyzed early-warning signals that are based on the estimation of the correlation properties of the system. These early-warning signals seem more robust in terms of asymmetry, but the change in the correlations is slower than the change in the statistical properties. It is important to notice that the in the regions far from criticality this system tends toward states in which either the correlations are nil (high temperatures) or trivial (low temperatures), but certainly this should not necessarily be the case for real complex systems which are not under our control. However, we believe that it is enlightening to understand how these early warnings behave under precisely controlled circumstances, specially because the main goal of early warning theory is to detect critical thresholds in systems where there is no a priori knowledge of whether a critical transition is near or even exists at all. Analyzing the correlations behavior by means of the Power Spectral Density in all the temporal scales available to the system lets us interpret the spectrum reddening as a characteristic way in which the scale invariance properties of critical systems disappear when the system departs from criticality. For systems poised at criticality, a power law in the PSD appears, and this power law is an indication that the system possesses  scale invariance in the time domain. When the system departs from criticality the PSD splits in two distinct regions. The low frequency regime is characterized by a flat spectrum, while the high frequency domain exhibits a power law. The crossover frequency that divides these two regions depends on how far the system is from the critical point. When the temperature is far away from the critical value, the spectrum flattens out completely, reflecting a total lack of correlations. At criticality, a single, full power law appears, i.e., the system acquires long term memory. We propose that the frequency at which this crossover occurs can be used as an early-warning signal that can indicate to what extent the temporal correlations have been lost, as well as provide a measure of the loss of temporal scale invariance in the system.


\section*{Acknowledgments}
This work was supported in part by grants from CONACyT-Mexico and the UNAM - DGAPA - PAPIIT IA100914 project.
            
\bibliography{ising8}

\begin{thebibliography}{10}
\providecommand{\url}[1]{\texttt{#1}}
\providecommand{\urlprefix}{URL }
\expandafter\ifx\csname urlstyle\endcsname\relax
  \providecommand{\doi}[1]{doi:\discretionary{}{}{}#1}\else
  \providecommand{\doi}{doi:\discretionary{}{}{}\begingroup
  \urlstyle{rm}\Url}\fi
\providecommand{\bibAnnoteFile}[1]{%
  \IfFileExists{#1}{\begin{quotation}\noindent\textsc{Key:} #1\\
  \textsc{Annotation:}\ \input{#1}\end{quotation}}{}}
\providecommand{\bibAnnote}[2]{%
  \begin{quotation}\noindent\textsc{Key:} #1\\
  \textsc{Annotation:}\ #2\end{quotation}}
\providecommand{\eprint}[2][]{\url{#2}}

\bibitem{Prokopenko:2009cj}
Prokopenko M, Boschetti F, Ryan AJ (2009) An information-theoretic primer on
  complexity, self-organization, and emergence.
\newblock Complexity 15: 11--28.
\bibAnnoteFile{Prokopenko:2009cj}

\bibitem{Gershenson:2012ft}
Gershenson C, Fern{\'a}ndez N (2012) Complexity and information: Measuring
  emergence, self-organization, and homeostasis at multiple scales.
\newblock Complexity 18: 29--44.
\bibAnnoteFile{Gershenson:2012ft}

\bibitem{Crutchfield:1989hp}
Crutchfield JP, Young K (1989) Inferring statistical complexity.
\newblock Phys Rev Lett 63: 105--108.
\bibAnnoteFile{Crutchfield:1989hp}

\bibitem{Razak-FA:2014uq}
Razak~FA JH (2014) Quantifying `causality' in complex systems: Understanding
  transfer entropy.
\newblock PLoS One 9(6): e99462.
\bibAnnoteFile{Razak-FA:2014uq}

\bibitem{Scheffer:2009sy}
Scheffer M, Bascompte J, Brock WA, Brovkin V, Carpenter SR, et~al. (2009)
  Early-warning signals for critical transitions.
\newblock Nature 461: 53--59.
\bibAnnoteFile{Scheffer:2009sy}

\bibitem{Sole:1996ng}
Sol{\'e} RV, Manrubia SC, Luque B, Delgado J, Bascompte J (1996) Phase
  transitions and complex systems: Simple, nonlinear models capture complex
  systems at the edge of chaos.
\newblock Complexity 1: 13--26.
\bibAnnoteFile{Sole:1996ng}

\bibitem{Bak:1988bx}
Bak P, Tang C, Wiesenfeld K (1988) Self-organized criticality.
\newblock Phys Rev A 38: 364--374.
\bibAnnoteFile{Bak:1988bx}

\bibitem{Landa:2011cz}
Landa E, Morales IO, Fossion R, Str\'ansk\'y P, Vel\'azquez V, et~al. (2011)
  Criticality and long-range correlations in time series in classical and
  quantum systems.
\newblock Phys Rev E 84: 016224.
\bibAnnoteFile{Landa:2011cz}

\bibitem{Kauffman:1993zw}
Kauffman S (1993) The origins of order : self-organization and selection in
  evolution.
\newblock New York: Oxford University Press.
\bibAnnoteFile{Kauffman:1993zw}

\bibitem{Carpenter:1999il}
Carpenter SR, Ludwig D, Brock WA (1999) Management of eutrophication for lakes
  subject to potentially irreversible change.
\newblock Ecological Applications 9: pp. 751-771.
\bibAnnoteFile{Carpenter:1999il}

\bibitem{deYoung:2008to}
deYoung B, Barange M, Beaugrand G, Harris R, Perry RI, et~al. (2008) Regime
  shifts in marine ecosystems: detection, prediction and management.
\newblock Trends in Ecology \& Evolution 23: 402--409.
\bibAnnoteFile{deYoung:2008to}

\bibitem{Rietkerk:2004qc}
Rietkerk M, Dekker SC, de~Ruiter PC, van~de Koppel J (2004) Self-organized
  patchiness and catastrophic shifts in ecosystems.
\newblock Science 305: 1926-1929.
\bibAnnoteFile{Rietkerk:2004qc}

\bibitem{Dai:2012tx}
Dai L, Vorselen D, Korolev KS, Gore J (2012) Generic indicators for loss of
  resilience before a tipping point leading to population collapse.
\newblock Science 336: 1175-1177.
\bibAnnoteFile{Dai:2012tx}

\bibitem{May:2008wu}
May RM, Levin SA, Sugihara G (2008) Complex systems: Ecology for bankers.
\newblock Nature 451: 893--895.
\bibAnnoteFile{May:2008wu}

\bibitem{Litt:2001wc}
Litt B, Esteller R, Echauz J, D'Alessandro M, Shor R, et~al. (2001) Epileptic
  seizures may begin hours in advance of clinical onset.
\newblock Neuron 30: 51--64.
\bibAnnoteFile{Litt:2001wc}

\bibitem{McSharry:2003bq}
McSharry PE, Smith LA, Tarassenko L (2003) Prediction of epileptic seizures:
  are nonlinear methods relevant?
\newblock Nat Med 9: 241--242.
\bibAnnoteFile{McSharry:2003bq}

\bibitem{Venegas:2005qp}
Venegas JG, Winkler T, Musch G, Vidal~Melo MF, Layfield D, et~al. (2005)
  Self-organized patchiness in asthma as a prelude to catastrophic shifts.
\newblock Nature 434: 777--782.
\bibAnnoteFile{Venegas:2005qp}

\bibitem{Kleinen:2003tk}
Kleinen T, Held H, Petschel-Held G (2003) The potential role of spectral
  properties in detecting thresholds in the earth system: application to the
  thermohaline circulation.
\newblock Ocean Dynamics 53: 53-63.
\bibAnnoteFile{Kleinen:2003tk}

\bibitem{Ising:1925qd}
Ising E (1925) Beitrag zur theorie des ferromagnetismus.
\newblock Z Physik 31: 253-258.
\bibAnnoteFile{Ising:1925qd}

\bibitem{Wang:1993hl}
Wang J (1993) Critical slowing down of the two-dimensional kinetic ising model
  with glauber dynamics.
\newblock Phys Rev B 47: 869--871.
\bibAnnoteFile{Wang:1993hl}

\bibitem{bib:Marinazzo_2014}
Marinazzo D, Pellicoro M, Wu G, Angelini L, Cort{\'e}s JM, et~al. (2014)
  Information transfer and criticality in the ising model on the human
  connectome.
\newblock PLoS ONE 9(4): e93616.
\bibAnnoteFile{bib:Marinazzo_2014}

\bibitem{bib:Das_2014}
Das TK, Abeyasinghe PM, Crone JS, et~al (2014) Highlighting the
  structure-function relationship of the brain with the ising model and graph
  theory.
\newblock Biomed Res Int : 237898.
\bibAnnoteFile{bib:Das_2014}

\bibitem{bib:Kitzbichler_2009}
Kitzbichler MG, Smith ML, Christensen SR, Bullmore E (2009) Broadband
  criticality of human brain network synchronization.
\newblock PLoS Comput Biol 5(3): e1000314.
\bibAnnoteFile{bib:Kitzbichler_2009}

\bibitem{bib:Torquato_2011}
Torquato S (2011) Toward an ising model of cancer and beyond.
\newblock Phys Biol 8(1): 015017.
\bibAnnoteFile{bib:Torquato_2011}

\bibitem{bib:Henry_2013}
Henry ER, Best RB, Eaton WA (2013) Comparing a simple theoretical model for
  protein folding with all-atom molecular dynamics simulations.
\newblock Proc Natl Acad Sci USA 110(44): 17880--17885.
\bibAnnoteFile{bib:Henry_2013}

\bibitem{bib:Liu_1993}
Liu Y, Dilger JP (1993) Application of the one- and two dimensional ising
  models to studies of cooperativity between ion channels.
\newblock Biophys J 64(1): 26--35.
\bibAnnoteFile{bib:Liu_1993}

\bibitem{bib:Majewski_2001}
Majewski J, Li H, Ott J (2001) The ising model in physics and statistical
  genetics.
\newblock Am J Hum Genet 69(4): 853--862.
\bibAnnoteFile{bib:Majewski_2001}

\bibitem{bib:Rice_2003}
Rice JJ, Stolovitzky G, Tu Y, Tombe PP (2003) Ising model of cardiac thin
  filament activation with nearest-neighbor cooperative interactions.
\newblock Biophys J 84: 897--909.
\bibAnnoteFile{bib:Rice_2003}

\bibitem{Castellano:2009}
Castellano C, Fortunato S, Loreto V (2009) Statistical physics of social
  dynamics.
\newblock Rev Mod Phys 81: 591- 646.
\bibAnnoteFile{Castellano:2009}

\bibitem{Metropolis:1953qr}
Metropolis (1953) Equation of state calculations by fast computing machines.
\newblock J Chem Phys 21: 1087-1092.
\bibAnnoteFile{Metropolis:1953qr}

\bibitem{Glauber:1963zf}
Glauber RJ (1963) Time‐dependent statistics of the ising model.
\newblock Journal of Math Phys 4: 294.
\bibAnnoteFile{Glauber:1963zf}

\bibitem{Dakos:2012ht}
Dakos V, Carpenter SR, Brock WA, Ellison AM, Guttal V, et~al. (2012) Methods
  for detecting early warnings of critical transitions in time series
  illustrated using simulated ecological data.
\newblock PLoS ONE 7(7): e41010.
\bibAnnoteFile{Dakos:2012ht}

\bibitem{Carpenter:2006hb}
Carpenter SR, Brock WA (2006) Rising variance: a leading indicator of
  ecological transition.
\newblock Ecology Letters 9: 311--318.
\bibAnnoteFile{Carpenter:2006hb}

\bibitem{Guttal:2008hw}
Guttal V, Jayaprakash C (2008) Changing skewness: an early warning signal of
  regime shifts in ecosystems.
\newblock Ecology Letters 11: 450--460.
\bibAnnoteFile{Guttal:2008hw}

\bibitem{Dakos:2008lp}
Dakos V, Scheffer M, van Nes EH, Brovkin V, Petoukhov V, et~al. (2008) Slowing
  down as an early warning signal for abrupt climate change.
\newblock Proceedings of the National Academy of Sciences 105: 14308-14312.
\bibAnnoteFile{Dakos:2008lp}

\bibitem{Peng:1994ye}
Peng CK, Buldyrev SV, Havlin S, Simons M, Stanley HE, et~al. (1994) Mosaic
  organization of dna nucleotides.
\newblock Phys Rev E 49: 1685--1689.
\bibAnnoteFile{Peng:1994ye}

\bibitem{Dakos:2011}
Dakos V, K{\'e}fi S, Rietkerk M, van Nes E, Scheffer M (2011) Slowing down in
  spatially patterned ecosystems at the brink of collapse.
\newblock The American Naturalist 177: E153-E166.
\bibAnnoteFile{Dakos:2011}

\bibitem{Kefi:2014}
K{\'e}fi S, Guttal V, Brock WA, Carpenter SR, Ellison AM, et~al. (2014) Early
  warning signals of ecological transitions: Methods for spatial patterns.
\newblock PLoS ONE 9(3): e92097.
\bibAnnoteFile{Kefi:2014}

\bibitem{jeldtoft:1998}
Jeldtoft JH (1998) Self-Organized Criticality: Emergent Complex Behavior in
  Physical and Biological Systems.
\newblock Cambridge University Press.
\bibAnnoteFile{jeldtoft:1998}

\bibitem{Champeney:1989}
Champeney DC (1989) A Handbook of Fourier Theorems.
\newblock Cambridge University Press.
\bibAnnoteFile{Champeney:1989}

\bibitem{Kobayashi:1982so}
Kobayashi M, Musha T (1982) 1/f fluctuation of heartbeat period.
\newblock Biomedical Engineering, IEEE Transactions on BME-29: 456-457.
\bibAnnoteFile{Kobayashi:1982so}

\bibitem{Appel:1989wk}
Appel ML, Berger RD, Saul J, Smith JM, Cohen RJ (1989) Beat to beat variability
  in cardiovascular variables: Noise or music?
\newblock Journal of the American College of Cardiology 14: 1139 - 1148.
\bibAnnoteFile{Appel:1989wk}

\bibitem{Iyengar-N:1996pd}
Iyengar N, Peng CK, Morin R, Goldberger A, Lipsitz L (1996) Age-related
  alterations in the fractal scaling of cardiac interbeat interval dynamics.
\newblock Am J Physiol 271: 1078-1084.
\bibAnnoteFile{Iyengar-N:1996pd}

\bibitem{Lipsitz:1990zg}
Lipsitz LA, Mietus J, Moody GB, Goldberger AL (1990) Spectral characteristics
  of heart rate variability before and during postural tilt. relations to aging
  and risk of syncope.
\newblock Circulation 81: 1803-10.
\bibAnnoteFile{Lipsitz:1990zg}

\end{thebibliography}

%
%
%

\section*{Figure Legends}
%

%
%
%

%
%

\end{document}